\documentclass{article}%
\usepackage{amsmath}
\usepackage{amssymb}
\usepackage{epsfig}
\usepackage{latexsym}
\usepackage{amsfonts}
\usepackage{graphicx}%
\begin{document}
\parindent 0mm \setlength{\parskip}{\baselineskip} \thispagestyle{empty}
\pagenumbering{arabic} \setcounter{page}{0} \mbox{ }
\rightline{UCT-TP-266/06}\newline\rightline{November 2006}\newline%
\vspace{0.2cm}
%\medskip

\begin{center}
{\Large \textbf{QCD vacuum condensates from tau-lepton decay data}}
{\LARGE \footnote{{\LARGE {\footnotesize Supported in part by DFG (Germany)
and NRF (South Africa) .}}}}
%\vspace{1cm}

\textbf{Cesareo A. Dominguez}$^{(a)}$, \textbf{Karl Schilcher$^{(b)}$%
}$\bigskip$

$^{(a)}$Centre for Theoretical Physics and Astrophysics\\[0pt]University of
Cape Town, Rondebosch 7700, South Africa

$^{(b)}$Institut f\"{u}r Physik, Johannes Gutenberg-Universit\"{a}t\\[0pt]%
Staudingerweg 7, D-55099 Mainz, Germany\newline
\vspace{0.5cm}

\textbf{Abstract}
\end{center}

\noindent The QCD vacuum condensates in the Operator Product Expansion are
extracted from the final ALEPH data on vector and axial-vector spectral
functions from $\tau$-decay. Weighted Finite Energy Sum Rules are employed in
the framework of both Fixed Order and Contour Improved Perturbation Theory. An
overall consistent picture satisfying chirality constraints can be achieved
only for values of the QCD scale below some critical value $\Lambda\simeq350
\; \mbox{MeV}$. For larger values of $\Lambda$, perturbation theory overwhelms
the power corrections. A strong correlation is then found between $\Lambda$
and the resulting values of the condensates. Reasonable accuracy is obtained
up to dimension $d=8$, beyond which no meaningful extraction is possible.


\newpage
\bigskip
\noindent
\section{Introduction}
\noindent
In the absence of exact analytical solutions to QCD at the Fermi scale, approximation methods are required to confront the theory with hadronic information. A particularly appealing method is that of QCD sum rules \cite{SVZ}, based on the Operator Product Expansion (OPE) of current correlators, and on the notion of quark-hadron duality. Dispersion relations relate hadronic parameters, such as masses and couplings, to universal QCD parameters, e.g. the strong coupling constant in the perturbative QCD sector,
and vacuum expectation values of gauge invariant quark and gluon field operators in the non-perturbative sector. These vacuum condensates are organized according to dimension, and their numerical values can be determined from data on hadronic spectral functions, e.g. as obtained from $e^{+}e^{-}$ annihilation, or $\tau$- lepton decay data. The former determines the vacuum condensates in the vector channel, while the latter allows for the extraction
of both vector and axial-vector channel condensates. Given the important role they play in applications of QCD sum rules, it is essential to determine the condensates with reasonable accuracy. Experimental data on hadronic spectral
functions reaches up to some maximum energy, beyond which one expects perturbative QCD (PQCD) to provide a reasonable description of the current correlator. This so called continuum threshold, $s_{0}$, is typically $s_{0}\simeq1-3\;\mbox{GeV}^{2}$. The vacuum condensates are then expressed in terms of a dispersive integral of the experimental spectral function up to $s=s_{0}$, followed by a calculable integral of the PQCD expression of the correlator. The latter obviously involves the strong coupling constant
$\alpha_{s}(s)$. Hence, there results a correlation between the extracted values of the vacuum condensates and the input value of the QCD scale $\Lambda$ entering $\alpha_{s}(s)$. The resulting sum rules based on Cauchy's Theorem are called Finite Energy Sum Rules (FESR). A valuable feature of this approach to the determination of the condensates is that, to first order in $\alpha_{s}(s)$, radiative corrections to the condensates do not
induce mixing of condensates of different dimension in a given FESR \cite{LAUNER}. This decoupling is absent in other sum rules, e.g. Laplace or Gaussian, which then complicates substantially the analysis.\\

Some of the early determinations of the condensates \cite{BERTL} used values of $\Lambda$ in the range $\Lambda\simeq100-200\;\mbox{MeV}$, which is substantially below the current range $\Lambda\simeq300-400\;\mbox{MeV}$
\cite{PDG}. Such high values of $\Lambda$ seriously complicate the task of extracting the condensates, as we pointed out a few times in the past \cite{DS1}-\cite{DS2}. The problem is that with increasing $\Lambda$, PQCD becomes the dominant contribution to the FESR, and overwhelms the power corrections above some critical value $\Lambda_{c}$. We have estimated \cite{DS1}-\cite{DS2} this to happen for $\Lambda_{c}\simeq 330-350\;\mbox{MeV}$. Another problem affecting the extraction of the vacuum condensates is the slow saturation of dispersive integrals by the hadronic spectral function data. Unlike the situation in deep inelastic electron-proton scattering, where precocious scaling is observed, the approach to asymptotia in tau-lepton decay does not appear to be as fast. The final ALEPH \cite{ALEPH2} data for
the chiral spectral function $v(s)-a(s)$ shows clearly that this  function has not yet reached its asymptotic form dictated by PQCD, i.e. it does not vanish, even at the highest energies attainable in $\tau $-decay. If the asymptotic regime had been reached precociously, let us say at $s \simeq 2\;\mbox{GeV}^{2}$, then it would have been straightforward to calculate the non-perturbative condensates with the help of the Cauchy integral. Since this is not the case, some method to improve convergence must be used. We have shown \cite{DS2}, \cite{DS3}-\cite{BORDES} that in the framework of FESR this can be done after suitably reducing the impact of the high energy region in the dispersive integral by using weighted sum rules. We used the data in a weighted linear
combination of the first two Weinberg sum rules, which follows from the  absence of  dimension $d=2$ and $d=4$ operators  in the chiral correlator, to demonstrate the precocious saturation of the FESR and the remarkable effectiveness of the method. Motivated by this success, we
determined a number of QCD condensates by making maximal use of the  absence of  $d=2$ and $d=4$ chiral operators, and requiring \textbf{strong stability}, i.e. we varied the radius $s_{0}$ in the Cauchy integral beginning at the end of $\tau$-decay phase space and required that the condensates calculated from the data should be reasonably constant for all
$s_{0}$, in some finite region, including the end of phase space \cite{BORDES}. The results are in agreement with most independent determinations \cite{CHIRAL}.

The next step is to proceed from chiral condensates to the extraction of individual vector and axial-vector condensates. In this task, it is important to achieve a consistent picture, e.g. the $d=4$ condensates should be chiral symmetric (neglecting the small term $m_{q} <\bar{q}q>$) as they are proportional to the gluon condensate. The same symmetry is expected of potential $d=2$ terms in the OPE, if present at all \cite{DS1}, \cite{CADC2}. In addition, individual values for vector and axial-vector condensates of
dimension $d =6$ should be consistent with independent determinations of chiral condensates of the same dimension \cite{BORDES}-\cite{CHIRAL}. This already poses a problem, as the chiral sum rules are independent of $\Lambda$ (for vanishing light quark masses), while the individual FESR are expected to introduce an appreciable correlation between $\Lambda$ and the condensates. The question is how strong is this correlation. In this context, an issue often overlooked is the manifest difference between the saturation of FESR by
data in the vector as opposed to the axial-vector channel. The former involves the rho-meson resonance peak at relatively low energy with a narrow width, while the latter has the very broad $a_{1}$ resonance above $1 \; \mbox{GeV}$.
As a result of this, the difference between the hadronic integral and the PQCD contribution, which essentially gives the condensates, is saturated differently in the two channels. For instance, for dimension $d = 2$,
saturation in the vector channel is from above, while in the axial-vector case it is from below. An important additional constraint is provided by data on the $\tau$ hadronic width $R_{\tau}$, which involves QCD perturbation theory
as well as vacuum condensates of dimension $d=6$ and $d=8$. The theoretical expression for $R_{\tau}$ involves a different integration kernel as the FESR giving the condensates; it can then be considered as a relatively independent constraint. Developments in the perturbative QCD sector, i.e. Contour Improved Perturbation Theory (CIPT)\cite{CIPT} should also be incorporated into the analysis. In fact, CIPT has been shown to provide a faster approach to asymptotia than Fixed Order Perturbation Theory (FOPT).\\

In this paper we carry out this self consistent analysis using the latest and final ALEPH data \cite{ALEPH2} on the separate vector and axial vector spectral functions. This data
differs from earlier ALEPH versions in that all data (as opposed to partial data sets) collected from 1991-1995 was used, and in that an increase in statistics  has resulted in a  reduction of the overall errors. It is important to mention at this point that using previous ALEPH data sets we were unable to obtain a consistent picture of the dimension $d=4$ condensates in the framework of FOPT. In fact, their signs differed, while they should have been identical. This problem is not present if the latest ALEPH data is used.
In Section 2 we use FOPT, and in Section 3 we determine
the condensates in the framework of CIPT. We favour a separate determination for each dimension and for each channel, for a particular value of $\Lambda$, as opposed to a single massive $\chi$-squared fit based on some exhaustive set
of sum rules. The latter poses the danger of masking the correlation between perturbative and non-perturbative contributions to the dispersive integrals. At the same time, it is very likely unstable with respect to the introduction of additional higher dimensional condensates. It can also overlook chirality constraints, necessary to achieve an overall consistent picture. As in the chiral case we find that a consistent picture can only be achieved on the basis of weighted, often called {\it pinched}, sum rules. An essential input to such sum rules is the absence of dimension $d=2$ condensates. We shall show that this is indeed a viable assumption. The conclusions are presented in Section 4.\\ 
\noindent
\section{Fixed Order Perturbation Theory}
\noindent We begin by defining the vector and axial-vector correlators

%Eq.2.1%
\begin{align}
\Pi_{\mu\nu}^{VV} (q^{2})  &  = i \; \int\; d^{4} \, x \; e^{i q x} \; \;
<0|T(V_{\mu}(x) \; \; V_{\nu}^{\dagger}(0))|0> ,\label{2.1}\\
&  = \; (- g_{\mu\nu} \; q^{2} + q_{\mu} q_{\nu}) \; \Pi_{V} (q^{2})
\;\nonumber
\end{align}

%Eq.2.2%
\begin{align}
\Pi_{\mu\nu}^{AA} (q^{2})  &  = i \; \int\; d^{4} \, x \; e^{i q x} \; \;
<0|T(A_{\mu}(x) \; \; A_{\nu}^{\dagger} (0) )|0> \;\label{2.2}\\
&  = \; (- g_{\mu\nu} \; q^{2} + q_{\mu} q_{\nu}) \; \Pi_{A} (q^{2}) - q_{\mu}
q_{\nu} \; \Pi_{0} (q^{2}) \;\nonumber
\end{align}

where $V_{\mu} = :(\bar{u} \gamma_{\mu} u-\bar{d} \gamma_{\mu} d):/2$, and
$A_{\mu} = :(\bar{u} \gamma_{\mu} \gamma_{5} u- \bar{d} \gamma_{\mu}
\gamma_{5} d):/2$. Considering these (charge neutral) currents implies, in
perturbative QCD, the normalization

%Eq.2.3%
\begin{equation}
\frac{1}{\pi}\operatorname{Im}\Pi_{V}^{QCD}\left(  s\right)  =\frac{1}{\pi
}\operatorname{Im}\Pi_{A}^{QCD}\left(  s\right)  =\frac{1}{8\pi^{2}}\left(
1+\frac{\alpha_{s}}{\pi} +...\right)  . \label{2.3}%
\end{equation}

The OPE of the current correlators Eqs.(\ref{2.1})-(\ref{2.2}) may be written as

%Eq.2.4%
\begin{equation}
8 \pi^{2} \Pi(Q^{2})|^{QCD}_{V,A}=\sum_{N= 0}^{\infty}\frac{1}{Q^{2N}}\;C_{2N}%
(Q^{2},\mu^{2})\;<O_{2N}(\mu^{2})>|_{V,A}\;, \label{2.4}%
\end{equation}

where the term $N=0$ stands for the purely perturbative QCD expression. An
alternative definition of the condensates, used often in the literature is

%Eq.2.5%
\begin{equation}
C_{2N}(Q^{2},\mu^{2})\;<O_{2N}(\mu^{2})>|_{V,A}\;\equiv8\pi^{2}\;\hat
{\mathcal{O}}_{2N}(Q^{2})|_{V,A}.\label{2.5}%
\end{equation}
\begin{figure}[h]
\begin{center}
\includegraphics[
height=4.0552in,
width=4.0552in
]{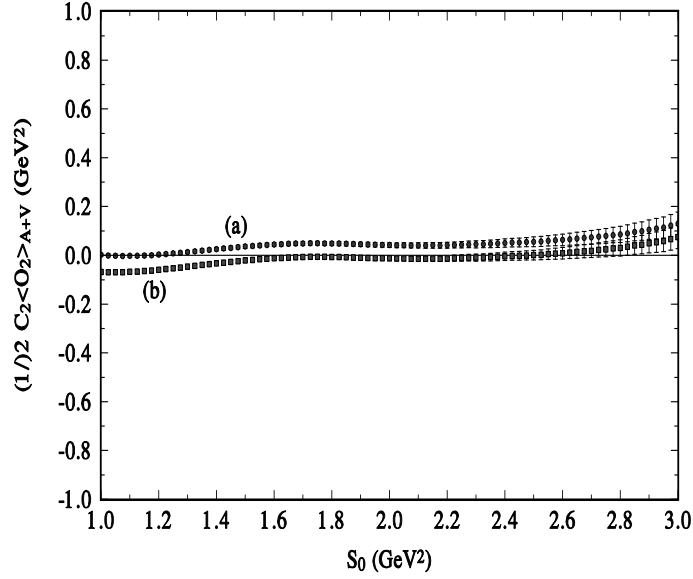}
\end{center}
\caption{The channel-averaged dimension $d=2$ vacuum condensate from
Eq.(\ref{2.7}) for (a): $\Lambda=300\;\mbox{MeV}$, and (b): $\Lambda
=350\;\mbox{MeV}$.}%
\label{C2O2AV}%
\end{figure}In FOPT one first calculates the Cauchy contour
integral for fixed $\alpha_{s}(\mu)$

%eq.2.6%
\begin{equation}
- \frac{1}{2 \pi i} \; \oint_{|s|=s_{0}} ds\; s^{N} \; \Pi(s)|^{QCD}_{V,A} \; =
\int_{0}^{s_{0}} ds \; s^{N} \;\frac{1}{\pi}\; Im\; \Pi(s)|_{V,A} \; ,
\label{2.6}%
\end{equation}

after which one performs the Renormalization Group (RG) improvement. This
leads to

%eq.2.7%
\begin{equation}
(-)^{N} C_{2N+2} < \hat{O}_{2N+2} > = 8 \pi^{2} \int_{0}^{s_{0}} ds \,s^{N}
\frac{1}{\pi} Im\; \Pi(s) \; - \frac{s_{0}^{N+1}}{(N+1)} \, I_{N}(s_{0})\,,
\label{2.7}%
\end{equation}

where $N=0,1,2,...$, the label V,A is omitted in the sequel, and $I_N(s_0)$ is given by the left hand side of Eq.(\ref{2.6}). An explicit calculation to five-loop order in PQCD yields
%eq.2.7b
%\begin{eqnarray}
%I_{N}(s_{0}) &\equiv& 8\pi %^{2}(N+1)\int_{0}^{s_0}\frac{ds}{s_{0}}%
%\left[ \frac{s}{s_0}\right] ^N \frac{1}{\pi } Im \; \Pi %_{QCD}(s) \nonumber \\ [.4cm]
%&=& -8 \pi
%^{2}(N+1)\frac{1}{2\pi i}\oint_{|s|=s_0} %\frac{ds}{s_0}\left[ 
%\frac{s}{s_0}\right] ^N \Pi _{QCD}(s) \;.\label{2.7b}
%\end{eqnarray}
%An explicit calculation to five-loop order yields
%eq.2.8

\begin{eqnarray}
I_N(s_0) &=& 1 \;+\; a_s(s_0) \;+\; [a_s(s_0)]^2\; \Bigl[ F_3 \;- \frac{\beta_1}{2} \frac{1}{(N+1)}\Bigr] \nonumber \\[.4cm]
&+& [a_s(s_0)]^3\; \Bigl[F_4 - \frac{1}{(N+1)} \;(F_3\, \beta_1 + \frac{\beta_2}{2})\,+ \, \frac{\beta_1^2}{2} \frac{1}{(N+1)^2}\Bigr] \nonumber \\[.4cm]
&+&[a_s(s_0)]^4\; \left\{k_3 - \frac{\pi^2}{4} \beta_1^2 F_3 - \frac{5}{24} \pi^2 \beta_1 \beta_2 -
\frac{1}{(N+1)} \Bigl[\frac{3}{2} \beta_1 ( F_4   \Bigr. \right. \nonumber \\ [.4cm]
&+& \Bigl. \left. \frac{\pi^2}{3} \frac{\beta_1^2}{4}) + \beta_2 F_3 + \frac{\beta_3}{2} - \frac{\pi^2}{8} \beta_1^3 \Bigr] + \frac{2}{(N+1)^2} \frac{\beta_1}{2}
(\frac{3}{2} \beta_1 F_3 + \frac{5}{4} \beta_2)
\right. \nonumber \\ [.4cm]
&-& \left. \frac{6}{(N+1)^3} \frac{\beta_1^3}{8} \right\} \; , \label{2.8}%
\end{eqnarray}

where for three flavours, as used throughout in this paper, and in the $\bar{MS}$ scheme, the coefficients are
\cite{BETA}: $\beta_{1} = - \frac{9}{2}$, $\beta_{2} = - \frac{67}{6}$,
$\beta_{3} = - \frac{3863}{192}$, $F_{3} = 1.6398$, $F_{4} = - 10.2839$. The
strong coupling constant in Eq.(\ref{2.8}), $a_s(s_0) \equiv \alpha_s(s_0)/\pi$, to five-loop order is \cite{CHET1}

%Eq.2.9
\begin{eqnarray}
\frac{\alpha^{(4)}_{s}(s_{0})}{\pi} &=&
\frac{\alpha^{(1)}_{s}(s_{0})}{\pi}
+ \Biggl (\frac{\alpha^{(1)}_{s}(s_{0})}{\pi}\Biggr )^{2}
\Biggl (\frac{- \beta_{2}}{\beta_{1}} {\rm ln} L \Biggr ) \nonumber \\ [.4cm]
&+& \Biggl (\frac{\alpha^{(1)}_{s}(s_{0})}{\pi}\Biggr )^{3} 
\Biggl (\frac{\beta_{2}^{2}}{\beta_{1}^{2}} ( {\rm ln}^{2} L -
{\rm ln} L -1) + \frac{ \beta_{3}}{\beta_{1}} \Biggr ) \nonumber \\ [.4cm]
&-& \Biggl (\frac{\alpha^{(1)}_{s}(s_{0})}{\pi}\Biggr )^{4}
\Biggl [\frac{\beta_{2}^3}{\beta_{1}^3} ({\rm ln}^3 L
-\frac{5}{2} {\rm ln}^{2} L - 2 {\rm ln} L + \frac{1}{2}) \nonumber \\ [.4cm]
&+& 3 \frac{\beta_2 \beta_3}{\beta_1^2} {\rm ln} L + \frac{b_3}{\beta_1}
\Biggr ] \; , \label{2.9}%
\end{eqnarray}

where
%Eq.2.10%
\begin{equation}
\frac{\alpha^{(1)}_{s}(s_{0})}{\pi} \equiv\frac{- 2}{\beta_{1} L}\; ,
\label{2.10}%
\end{equation}

with $L \equiv\mathrm{ln} (s_{0}/\Lambda^{2})$, and

%Eq.2.11
\begin{eqnarray}
b_3&=&\frac{1}{4^4} \Biggl[ \frac{149753}{6} + 3564 \zeta_3
-(\frac{1078361}{162} + \frac{6508}{27} \zeta_3 ) n_F \nonumber \\
&+& (\frac{50065}{162} + \frac{6472}{81} \zeta_3 ) n_F^2 +
 \frac{1093}{729}
n_F^3 \Biggr ] \; , \label{2.11}%
\end{eqnarray}

with $\zeta_{3} = 1.202$. The five-loop constant $k_{3}$ is not yet known. We
have estimated it \cite{DS1} assuming a geometric series behaviour for those
constants not determined by the renormalization group, i.e. $k_{3} \simeq
k_{2}^{2}/k_{1} \simeq25$, with $k_{1} = F_{3}$ and $k_{2} = F_{4} + \pi^{2}
\beta_{1}^{2}/12$. This is in good qualitative agreement with other estimates
\cite{K3}.

In order to compute the hadronic integral in Eq.(7) we have used the latest and final ALEPH data set \cite{ALEPH2}. This set includes together with each data point the statistical error, and also the full error correlation matrix. The integration interval is divided in bins, and the hadronic integral is computed as a function of $s_0$ in the standard fashion by addition, taking into account error propagation and correlation.
From Eq.(\ref{2.7}) with $N=0$,  we determined the dimension $d=2$ condensate in the vector and the axial-vector channel, for $\Lambda= 300 - 350 \;\mbox{MeV}$. The results do not yield a consistent picture, as the vector channel condensate is small and negative in the
stability region around $s_{0} \simeq1.7 - 2.3 \;\mbox{GeV}^{2}$, while in the axial-vector channel it is mostly positive and numerically much larger. However, constraining this condensate to be chiral-symmetric, and taking the average in the two channels leads to a very stable result, consistent with zero, as shown in Fig. 1. The criterion of stability, which is standard in this kind of analysis, is quite simple. Quantities such as the vacuum condensates should be  indepenedent of $s_0$. Therefore, their values are read in the region where they show the least dependence on $s_0$, which should be high enough to allow for the onset of perturbative QCD. Numerically the result is

%eq.2.12%
\begin{equation}
\frac{1}{2} \Bigl[C_{2}<\hat{O}_{2}>|_{V} + C_{2}<\hat{O}_{2}>|_{A} \Bigr] =
0.05 \pm0.05 \;\mbox{GeV}^{2} \;. \label{2.12}%
\end{equation}
\begin{figure}[h]
\begin{center}
\includegraphics[
height=4.0552in,
width=4.0552in
]{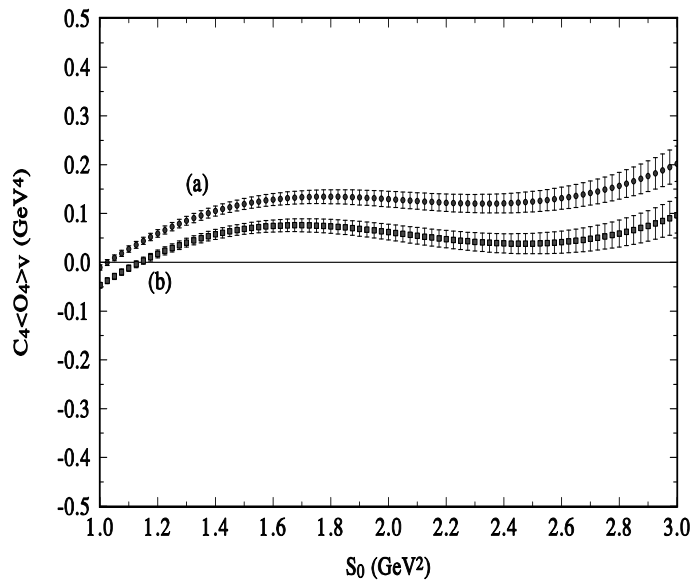}
\end{center}
\caption{The dimension $d=4$ vacuum condensate from the vector channel in CIPT, Eq.(\ref{3.6}), for (a): $\Lambda= 300\; \mbox{MeV}$, and (b): $\Lambda=350 \;\mbox{MeV}$.}%
\label{C4O4V}%
\end{figure}

\begin{figure}[h]
\begin{center}
\includegraphics[
height=4.0552in,
width=4.0552in
]{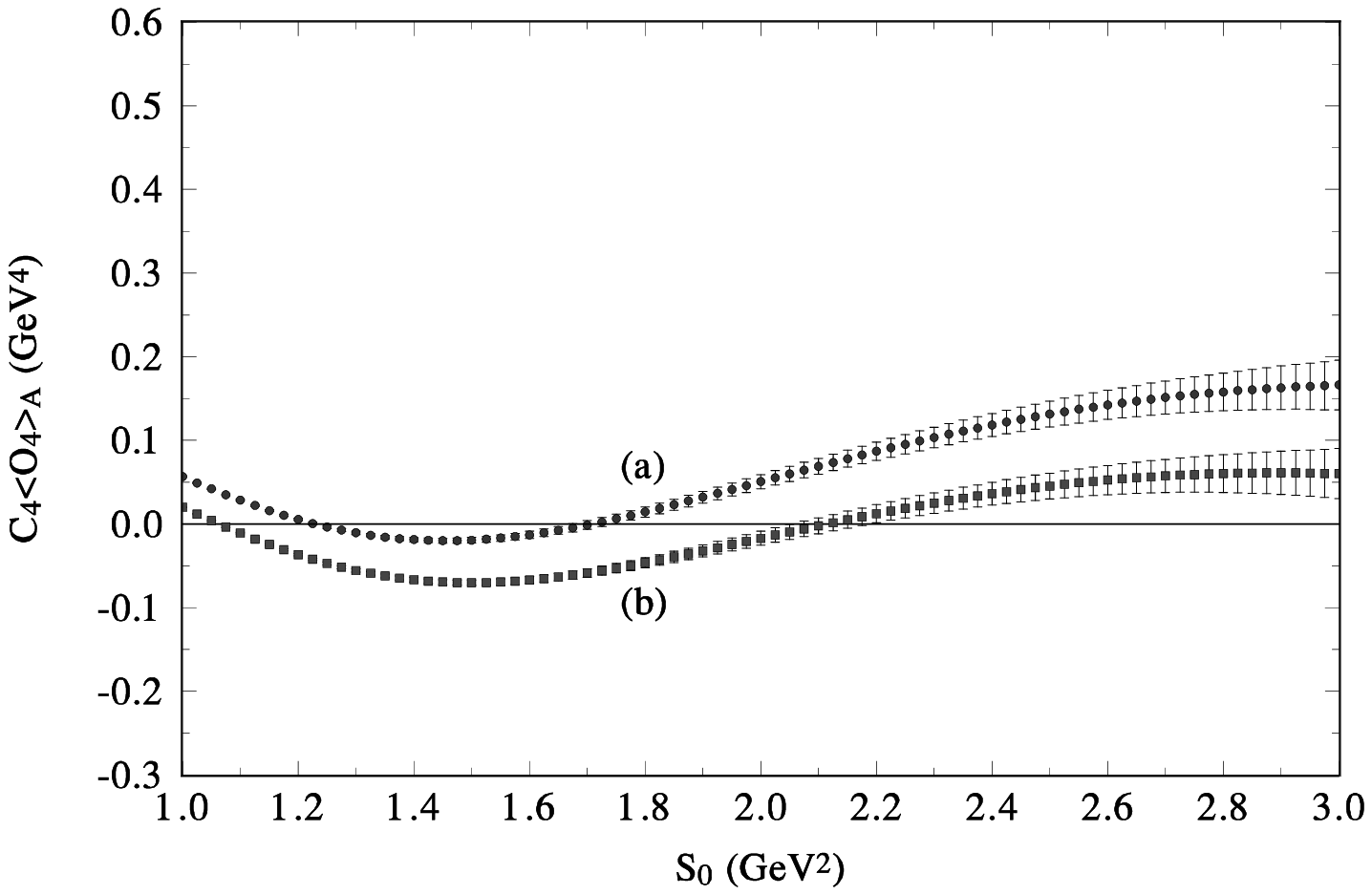}
\end{center}
\caption{The dimension $d=4$ vacuum condensate from the axial-vector channel in CIPT, Eq.(\ref{3.6}), for (a): $\Lambda=300\;\mbox{MeV}$, and(b):$\Lambda=350\;\mbox{MeV}$.}% 
\label{C4O4A}%
\end{figure}For higher dimensional condensates, $d\geq4$, results from these standard FESR are inconclusive. For instance, the $d=4$ gluon condensate is not positive definite in either channel. This is an indication that the asymptotic QCD regime has not been reached at the scales accessible in $\tau $-decay, at least in the vicinity of the positive real axis of the contour integral. An attempt has been made to improve this situation by introducing an integration kernel in Eq.(\ref{2.6}) that vanishes at $s=s_{0}$, i.e. one
makes the substitution $s^{N}\rightarrow\lbrack1-(s/s_{0})^{N}]$ for $N=1,2,...$. A meaningful use of these so called \textit{pinched moments
}relies heavily on the assumption that there is no dimension $ d =2$ condensate. Results from these weighted FESR are much better, provided $\Lambda \simeq300\;\mbox{MeV}$. For larger values of $\Lambda$ the condensates are consistent with zero. Numerically we obtain

%eq.2.13%
\begin{equation}
C_{4}<O_{4}>|_{V,A} = 0.10 \pm0.05 \; \mbox{GeV}^{4}\;\;\; (\Lambda= \; 300\;
\mbox{MeV})\; , \label{2.13}%
\end{equation}

%eq.2.14%
\begin{equation}
C_{6}<O_{6}>|_{V} = - 0.5 \pm0.2 \; \mbox{GeV}^{6}\;\;\;\;\;\;\;\;(\Lambda= \;
300 \;\mbox{MeV})\; , \label{2.14}%
\end{equation}

%eq.2.15%
\begin{equation}
C_{6}<O_{6}>|_{A} = - 0.3 \pm0.2 \; \mbox{GeV}^{6} \;\;\;\;\;\;\;\; (\Lambda=
\; 300 \;\mbox{MeV})\;. \label{2.15}%
\end{equation}

In the case of $d=8$, only the vector channel condensate shows reasonable stability and has the value

%eq.2.16%
\begin{equation}
C_{8}<O_{8}>|_{V} = 0.6 \pm0.2 \; \mbox{GeV}^{8}\;\;\;\;\;\; (\Lambda= \; 300
\;\mbox{MeV}) \;. \label{2.16}%
\end{equation}

The graphical dependence of the condensates on the continuum threshold $s_{0}$, being qualitatively similar in FOPT as in CIPT, will be shown in the next section. There, we discuss the use of CIPT together with \textit{pinched moments} which does lead to a more consistent determination of the first few vacuum condensates with somewhat improved accuracy.

\noindent

\section{Contour Improved Perturbation Theory}
\noindent Unlike the case of FOPT, where $\alpha_{s}(s_{0})$ is frozen in Cauchy's contour integral, and the RG is implemented after integration, in CIPT $\alpha_{s}$ is running and the RG is used before integrating. It has been shown\cite{CIPT} that in $\tau$-decay CIPT is superior to FOPT. To proceed we introduce the Adler function

%eq.3.1%
\begin{equation}
D(s) \equiv-s \frac{d}{ds} \Pi(s) \; . \label{3.1}%
\end{equation}

From Cauchy's theorem, and after an integration by parts, a relation between the contour integrals of $\Pi(s)$ and $D(s)$ is easily obtained

%eq.3.2%
\begin{equation}
\oint_{|s|=s_{0}} ds\; (\frac{s}{s_{0}})^{N} \;\Pi(s) = \frac{1}{N+1}
\;\frac{1}{s_{0}^{N}} \;\oint_{|s|=s_{0}} \frac{ds}{s}\; (s^{N+1} -
s_{0}^{N+1})\; D(s) \; . \label{3.2}%
\end{equation}
\begin{figure}[h]
\begin{center}
\includegraphics[
height=4.0552in,
width=4.0552in
]{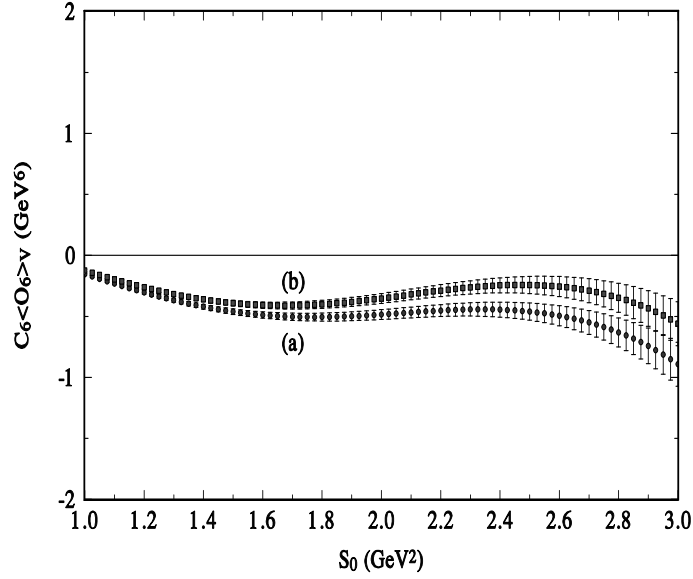}
\end{center}
\caption{The dimension $d=6$ vacuum condensate from the vector channel in
CIPT, Eq.(\ref{3.6}), for (a): $\Lambda= 300\; \mbox{MeV}$, and (b):
$\Lambda=350 \;\mbox{MeV}$.}%
\label{C6O6V}%
\end{figure}
After RG improvement, the perturbative expansion of the Adler
function can be written as

%eq.3.3%
\begin{equation}
D(s) = \frac{1}{8\; \pi^{2}} \sum_{m = 0} K_{m} \; \Bigl[\frac{\alpha_{s}%
(-s)}{\pi}\Bigr]^{m} \; , \label{3.3}%
\end{equation}

where \cite{CHET2} $K_{0} = K_{1} =1$, $K_{2} \equiv F_{3} = 1.6398$ , $K_{3} = F_{4} + \frac{\pi^{2}}{12} \beta_{1}^{2} = 6.3710,$, for three flavours, and the unknown coefficient $K_{4} \equiv k_{3} \simeq25$. Next, we introduce the moments

%eq.3.4%
\begin{equation}
M_{N}(s_{0}) = \frac{1}{2 \pi} \frac{1}{(N+1)} \sum_{m=0} K_{m} \;
[I_{N+1,m}(s_{0}) - I_{0,m}(s_{0})] \;, \label{3.4}%
\end{equation}

where

%eq.3.5%
\begin{equation}
I_{N,m} \equiv i \; \oint_{|s|=s_{0}} ds\; (\frac{s}{s_{0}})^{N} \;
[\frac{\alpha_{s}(-s)}{\pi}]^{m} \; . \label{3.5}%
\end{equation}
\begin{figure}[h]
\begin{center}
\includegraphics[
height=4.0552in,
width=4.0552in
]{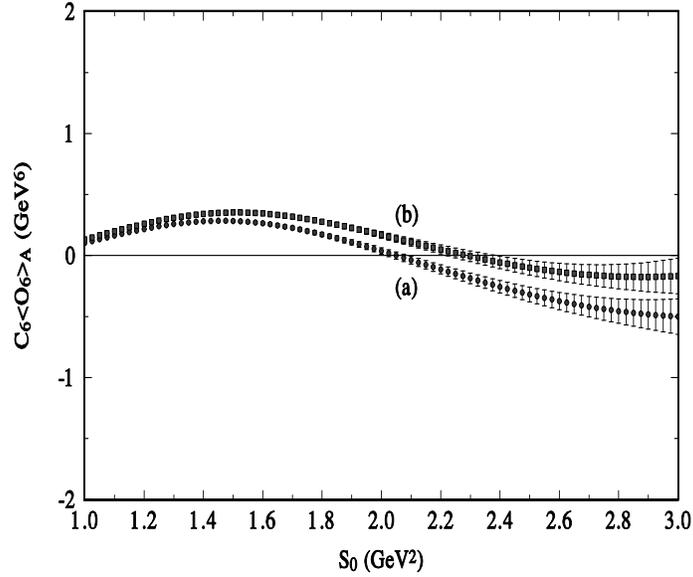}
\end{center}
\caption{The dimension $d=6$ vacuum condensate from the axial-vector channel in CIPT, Eq.(\ref{3.6}), for (a): $\Lambda= 300\; \mbox{MeV}$, and (b): $\Lambda=350 \;\mbox{MeV}$.}%
\label{C6O6A}%
\end{figure}The vacuum condensates can then be expressed as

%eq.3.6%
\begin{align}
C_{2N+2} < \hat{O}_{2N+2} >  &  = (-)^{N+1}\; 8 \pi^{2}\; s_{0}^{N} \;
\int_{0}^{s_{0}} \; ds \left[  1 - (\frac{s}{s_{0}})^{N} \right]  \frac{1}%
{\pi} Im \; \Pi(s)\nonumber\\[.4cm]
&  (-)^{N} s_{0}^{N+1} [ M_{0}(s_{0}) - M_{N}(s_{0})] \;, \label{3.6}%
\end{align}

where a \textit{pinched} integration kernel has been introduced. The l.h.s. of Eq.(\ref{3.6}) includes radiative corrections up to order $\cal{O}$$(\alpha_{s})$. Mixing of operators of dimension lower or higher than the leading term occurs only at order $\cal{O}$$(\alpha_{s}^2)$ \cite{LAUNER}, which can be safely neglected. In the framework of CIPT \cite{CIPT} the contour integrals, Eq. (\ref{3.5}), are calculated numerically using for $\alpha_{s}$ a numerical solution to the RG equation

%eq.3.7%
\begin{equation}
\mu\; \frac{d a_{s}(\mu^{2})}{d \mu} = \sum_{n=1} \beta_{n} [a_{s}(\mu
^{2})]^{n+1} \;. \label{3.7}%
\end{equation}

We have used a Modified Euler method to solve the RG equation at each integration point in the complex s-plane, and a single step numerical integration to compute the contour integral.

Using Eq.(\ref{3.6}) we first determine the dimension $d=2$ vacuum condensates in the vector and in the axial-vector channel. Results are comparable to those from FOPT, so the numerical value of the channel-average condensate is still
that of Eq.(\ref{2.12}). Next, for $d=4$, and assuming  $C_{2}<O_{2}>|_{V,A} = 0$, the results in the vector and the axial-vector channel are shown in Figs.2 and 3, respectively, for the two values $\Lambda=
300 \; \mbox{MeV}$, curve (a), and $\Lambda= 350 \; \mbox{MeV}$, curve (b). Although results in the axial-vector channel become consistent with those in the vector channel only close to the end-point, at least such a consistency is
achieved. Numerically we find

%eq.3.8%
\begin{equation}
C_{4}<O_{4}>|_{V,A} = \left\{
\begin{array}
[c]{ll}%
0.15 \pm0.04\; \mbox{GeV}^{4} \;\;\;(\Lambda= 300\; \mbox{MeV}) & \\
0.07 \pm0.02 \; \mbox{GeV}^{4} \;\;\;(\Lambda= 350 \; \mbox{MeV}) &
\end{array}
\right.  \;. \label{3.8}%
\end{equation}

\begin{figure}[h]
\begin{center}
\includegraphics[
height=4.0552in,
width=4.0552in
]{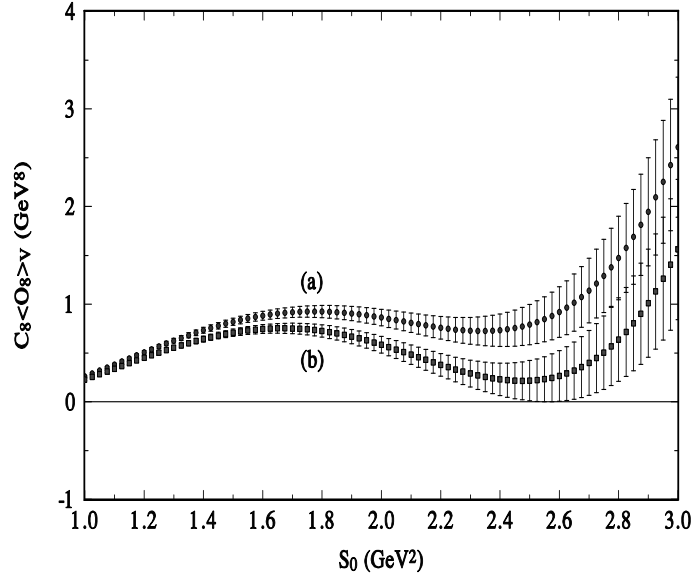}
\end{center}
\caption{The dimension $d=8$ vacuum condensate from the vector channel in
CIPT, Eq.(\ref{3.6}), for (a): $\Lambda= 300\; \mbox{MeV}$, and (b):
$\Lambda=350 \;\mbox{MeV}$.}%
\label{C8O8V}%
\end{figure}

For larger values of $\Lambda$ the process breaks down as $C_{4}<O_{4}>|_{V}$ vanishes and $C_{4}<O_{4}>|_{A}$ becomes negative. This confirms our earlier claims about the existence of a critical value $\Lambda_{c} \simeq330 - 350 \;
\mbox{MeV}$ beyond which no meaningful determination of the condensates is possible \cite{DS1}-\cite{DS2}. Proceeding to dimension $d=6$, results are shown in Figs. 4 and 5, for the same two choices of $\Lambda$. The condensate in the vector channel is reasonably stable giving the values

%eq.3.9%
\begin{equation}
C_{6}<O_{6}>|_{V} = \left\{
\begin{array}
[c]{ll}%
- 0.63 \pm0.12\; \mbox{GeV}^{6} \;\;\;(\Lambda= 300\; \mbox{MeV}) & \\
- 0.35 \pm0.12 \; \mbox{GeV}^{6} \;\;\;(\Lambda= 350 \; \mbox{MeV}) &
\end{array}
\right.  \; . \label{3.9}%
\end{equation}

Results in the axial-vector channel change sign at $s_{0} \simeq2.1 - 2.2 \; \mbox{GeV}^{2}$, and are then barely conclusive. However, we can use the reasonably accurate determination \cite{BORDES} of the chiral condensate of
$d=6$

%eq.3.10%
\begin{equation}
C_{6}<O_{6}>|_{V-A} = - 0.18 \pm0.04 \;, \label{3.10}%
\end{equation}
together with Eq. (\ref{3.9}) to obtain

%eq.3.11%
\begin{equation}
C_{6}<O_{6}>|_{A} = \left\{
\begin{array}
[c]{ll}%
- 0.45 \pm0.13\; \mbox{GeV}^{6} \;\;\;(\Lambda= 300\; \mbox{MeV}) & \\
- 0.17 \pm0.12 \; \mbox{GeV}^{6} \;\;\;(\Lambda= 350 \; \mbox{MeV}) &
\end{array}
\right.  \; . \label{3.11}%
\end{equation}

These results are in remarkable agreement with the values of $C_{6} <O_{6}>|_{A}$, if one were to read off the condensate close to the end point
(see Fig. 5)

%eq.3.12%
\begin{equation}
C_{6}<O_{6}>|_{A} = \left\{
\begin{array}
[c]{ll}%
- 0.50 \pm0.15\; \mbox{GeV}^{6} \;\;\;(\Lambda= 300\; \mbox{MeV}) & \\
- 0.18 \pm0.13 \; \mbox{GeV}^{6} \;\;\;(\Lambda= 350 \; \mbox{MeV}) &
\end{array}
\right.  \; . \label{3.12}%
\end{equation}

However, it should be kept in mind that the result Eq. (\ref{3.10}) is independent of $\Lambda$, as there is no perturbative contribution to chiral sum rules (for vanishing light quark masses). The sign of $C_{6}<O_{6}>|_{A}$ as obtained above does not agree with the so called \textit{Vacuum Saturation} (V.S.) approximation to estimate the four-quark contribution to $C_{6}<O_{6}>$. In fact, according to V.S. the ratio of vector to axial-vector condensates is negative

%eq.3.13%
\begin{equation}
\frac{C_{6}<O_{6}>|_{V}}{C_{6}<O_{6}>|_{A}} = - \frac{7}{11} \;. \label{3.13}%
\end{equation}

Reading off $C_{6}<O_{6}>|_{A}$ at lower values of $s_{0}$, where it is positive, would give $C_{6}<O_{6}>|_{A} \simeq0.2 - 0.3 \; \mbox{GeV}^{6}$,
leading to a chiral value $- (0.65 - 0.85) \; \mbox{GeV}^{6}$ in serious disagreement with Eq.(\ref{3.10}). From Fig.5 one could always speculate that the fully asymptotic region may have not been reached at $3\; \mbox{GeV}^2$, and that a sign change in that region could take place. It should be mentioned that using the previous versions of the ALEPH data base, we obtain a positive sign for $C_{6}<O_{6}>|_{A}$, and a negative sign for $C_{6}<O_{6}>|_{V}$, using FOPT and a pinched moment. However, this data base leads to different signs for $C_{4}<O_{4}>$ in the vector and axial-vector channels. \\

Not surprisingly, the situation with the dimension $d=8$ condensate is considerably worse. Results in the axial-vector channel are totally inconclusive: the condensate changes sign in the middle of the integration region for both values of $\Lambda$. For instance, for $s_{0} \geq1.8 \; \mbox{GeV}^{2}$, $C_{8}<O_{8}>|_{A}$ rises monotonically from zero to about $2.5 \; \mbox{GeV}^{8}$ for $\Lambda= 300 \;\; \mbox{MeV}$. The vector channel provides an acceptable result only for $\Lambda= 300 \; \mbox{MeV}$, as shown
in Fig. 6. Reading values in the stability region $s_{0} \simeq1.5 - 2.4 \; \mbox{GeV}^{2}$ gives

%eq.3.14%
\begin{equation}
C_{8}<O_{8}>|_{V} = 0.8 \pm0.2\; \mbox{GeV}^{8} \;\;\;(\Lambda= 300\;
\mbox{MeV}) \; . \label{3.14}%
\end{equation}

In this case it is not possible to use this result, together with the independent determination of the  condensate $C_{8}<O_{8}>|_{V-A}$ in the chiral limit, in order to obtain $C_{8}<O_{8}>|_{A}$. In fact, at $d=8$ there are chiral symmetric contributions, proportional to the gluon condensate and to the single quark condensate, which vanish in the chiral condensate \cite{BROAD}.
\begin{figure}[h]
\begin{center}
\includegraphics[
height=4.0552in,
width=4.0552in
]{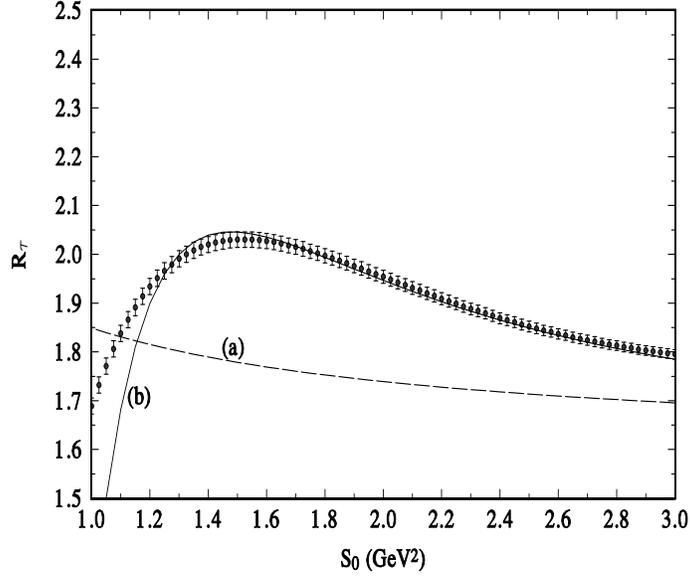}
\end{center}
\caption{The ratio $R_{\tau}$ in the vector channel. Data points are from
Eq.(\ref{3.15}). Theoretical curves are from CIPT, Eq.(\ref{3.16}), for
$\Lambda= 300\; \mbox{MeV}$, and (a):$\;\;\;$ $C_{6}<O_{6}> = C_{8}<O_{8}> =
0$, (b): $C_{6}<O_{6}>|_{V} = - 0.45 \;\mbox{GeV}^{6},\;$ and $\;\;\;\;
C_{8}<O_{8}>|_{V} = 0.78 \;\mbox{GeV}^{8}$.}%
\label{RTAUV300}%
\end{figure}

We finally turn to the $\tau$-ratio in either the vector or the axial-vector channel

%Eq.3.15%
\begin{equation}
R_{\tau}= 24 \pi^{2} |V_{ud}|^{2} S_{EW} \int_{0}^{s_{0}} \frac{ds}{s_{0}}
\Biggl [ 1 - 3 (\frac{s}{s_{0}})^{2} + 2 (\frac{s}{s_{0}})^{3} \Biggr ] \frac
{1}{\pi} Im \, \Pi(s) \;. \label{3.15}%
\end{equation}

The theoretical expression of $R_{\tau}$ which follows from Eqs.(\ref{2.4}) and (\ref{3.4}), again for either the vector or the axial-vector channel, is

%eq.3.16%
\begin{align}
R_{\tau}  &  = 3 |V_{ud}|^{2} S_{EW} \Bigl \{ M_{0}(s_{0}) - 3 \Bigl [ M_{2}%
(s_{0}) + \frac{C_{6}<O_{6}>}{s_{0}^{3}} \Bigr ]\nonumber\\[.4cm]
&  + 2 \Bigl [ M_{3}(s_{0}) - \frac{C_{8}<O_{8}>}{s_{0}^{4}}
\Bigr ] \Bigr \} \;. \label{3.16}%
\end{align}

The results for $R_{\tau}$ in the vector channel are shown in
Fig.\ref{RTAUV300}. Curve (a) corresponds to Eq.(\ref{3.16}) for $C_{6}<O_{6}> = C_{8}<O_{8}> =0$, and curve (b) for $C_{6}<O_{6}> = - 0.45 \; \mbox{GeV}^{6}$, $C_{8}<O_{8}> = 0.78 \; \mbox{GeV}^{8}$. The latter values  are in very good agreement with the results Eqs.(\ref{3.9}) and (\ref{3.14}). This is an independent confirmation of the strong correlation between the values of the condensates and the value of $\Lambda$. The situation in the axial-vector channel is inconclusive partly
because of the large uncertainties in the condensates, and partly because the data approach the asymptotic regime from below, rather from above as in the vector channel.

\section{Conclusion}

In the framework of FOPT, and using standard FESR without the \textit{pinched kernel} $[1 - (s/s_{0})^{N}]$, the dimension $d=2$ condensate is consistent with zero after requiring it to be chiral symmetric, and averaging over the two channels. However, the result, Eq.(\ref{2.12}), does not exclude a very
small value for this condensate. For dimension $d \geq4$ no self consistent picture is achieved. For instance, for $d=4$ the gluon condensate comes out with different signs in the vector and axial-vector channels. This overall situation improves with the inclusion of the \textit{pinched kernel} in the FESR, but then only if $\Lambda\simeq300 \; \mbox{MeV}$. Larger values of $\Lambda$ yield condensates consistent with zero. This agrees with our earlier observations about the existence of a critical value $\Lambda_{c} \simeq330 -
350 \; \mbox{MeV}$ beyond which no meaningful determination of the condensates is possible \cite{DS1}-\cite{DS2}. In the framework of CIPT, and using weighted FESR, the overall picture improves. It is possible to accommodate larger values of $\Lambda$ up to $\Lambda\simeq350 \; \mbox{MeV}$, and to
determine the condensates in the vector and axial-vector channels with reasonable accuracy. It is also possible to achieve consistency in terms of chirality: the gluon condensate turns out to have the same sign in both
channels, and the difference between the $d=6$ condensates is consistent with independent determinations from chiral sum rules. The sign of $C_{6} <O_{6}>|_{A}$, though, disagrees with the \textit{Vacuum Saturation} approximation, provided $C_{6}<O_{6}>|_{V}$ is negative, as we have found here. Of course, a change of sign in $C_{6} <O_{6}>|_{A}$ beyond $3\; \mbox{GeV}^2$ cannot be excluded. 
For dimension $d=8$, only the vector condensate can be determined unambiguously, and then only for $\Lambda= 300 \; \mbox{MeV}$. Given the strong correlation between $\Lambda$ and the condensates, we wish to suggest that in applications of QCD sum rules, values for the latter be used in
conjunction with the corresponding values of $\Lambda$. Because of this correlation, we find little merit in averaging the results obtained from this analysis at different values of $\Lambda$.

\end{document}